\documentclass[]{article}
\usepackage[cm]{fullpage}
\usepackage{url,color,graphicx,wrapfig,comment,subcaption}

\begin{document}

\newcommand{\ie}[1]{\emph{i.e.~}}

\newcommand{\red}[1]{\textcolor{red}{#1}}

\title{Identifying and modelling cognitive biases in mobility choices}
\author{Chloe CONRAD\\M1 student, Univ. Lyon 1 \and Carole ADAM\\Univ. Grenoble-Alpes - LIG}
\date{M1 internship report - October 2022 to June 2023}

\maketitle

\begin{abstract}
This report presents results from an M1 internship dedicated to agent-based modelling and simulation of daily mobility choices. This simulation is intended to be realistic enough to serve as a basis for a serious game about the mobility transition. In order to ensure this level of realism, we conducted a survey to measure if real mobility choices are made rationally, or how biased they are. Results analysed here show that various biases could play a role in decisions. We then propose an implementation in a GAMA agent-based simulation. 
\end{abstract}

\section{Introduction}

One of the central axes in the development of a more ecological society is the transition towards more sustainable modes of mobility. To trigger and successfully complete this transition, several tools and levers can be used. One of these tools is the implementation of public policies favoring the use of soft mobility modes \cite{ref7-huyghe}. However, data shows that this approach alone is not sufficient\footnote{Chiffres clés des transports 2023. Données et études statistiques pour le changement	climatique,	l’énergie, l’environnement, le	logement et	les	transports : \url{https://www.statistiques.developpement-durable.gouv.fr/chiffres-cles-des-transports-edition2023?rubrique=56dossier=1337}}. Another complementary approach consists in developing educational tools to facilitate this transition.

The Switch project\footnote{SwITCh - Simulating the transition of transport Infrastructures Toward smart and sustainable Cities: \url{https://www6.inrae.fr/switch}} follows this second approach, with the objective of developing a serious game to raise awareness about the mobility transition. The main idea of the project is to use game dynamics on top of an agent-based simulation, to trigger reflection in the player about their daily mobility choices. An agent-based simulation is a computer program in which autonomous intelligent entities called agents interact together and with their environment to carry out a certain task. These systems fall into the field of distributed artificial intelligence and are particularly useful for such educational purposes. Indeed, they allow the user to interact with the system, try out different scenarios, and control relevant parameters in order to understand their impact on the observed outputs. 

This report is structured as follows: Section~\ref{sec:soa} provides some background and details the previous work on which the current internship project is grounded; Section~\ref{sec:survey} describes our survey and analyses its results; Section~\ref{sec:simul} introduces our agent-based simulator, how it uses the survey results, and two experiments; Section~\ref{sec:cci} concludes the report.

\section{Background and previous work} \label{sec:soa}  

\subsection{A model of mobility choice}

The first mobility choice simulator in the SWITCH project was proposed by \cite{jacquier2021choice}. It defined a rational, multi-criteria decision process based on six criteria identified from sociological studies: comfort, speed, safety, cost, praticity, and ecology. Each individual agent $i$ has their own priorities for each criterion $c$, denoted $prio_i(c)$, showing how important this criterion is for the agent. Four mobility modes are studied: individual car, public transport (bus), bicycle, walking. Each mode $m$ has a value on each criteria $c$, denoted $val(m,c)$; this 'objective' value depends on the current city infrastructure and represents how well the mode $m$ satisfies the criterion $c$. Based on these heterogeneous priorities, and homogeneous values, each individual $i$ computes their score $score_i(m)$ for each mobility mode $m$, with the following formula:
$$score_i(m) = \sum_{c \in crits} val(m,c) * prio_i(c)$$

The rational choice for the agent $i$ is then the mobility mode that receives the maximal score. This model is interesting but was shown to be insufficient, in particular due to ignoring the role of habits \cite{brette2014reconsidering} that create an inertia in our mobility choices. The model was therefore later enriched with habits \cite{jacquier2021gamadays,isaga-habits} as well as cognitive biases \cite{isaga-bias}.

\subsection{Cognitive biases}

Cognitive biases are heuristics used by our cognitive system to facilitate decision-making \cite{tversky1974judgment}. They allow rapid reasoning during stressful or complex situations despite the incompleteness or uncertainty of the information necessary for rational decision-making. Although they are essential to our proper functioning, they can sometimes lead to irrational decision-making. Therefore they must be taken into account when constructing realistic human behaviour models.

The literature in cognitive science has identified a large number of cognitive biases \cite{ref4-ehrlinger, ref6-hilbert}, but not all of them are interesting to take into account in the context of daily mobility choice. Thus, before developing a simulation, it is necessary to identify those which play a role in our context and which may therefore be relevant to implement in our simulation. An initial study based on semi-structured interviews and bibliographic research was previously carried out and allowed us to select 3 biases:
\begin{itemize}
    \item Confirmation bias: tendency to only consider information that confirms our opinions and choices.
    \item Reactance bias: negative reaction to an external attempt to influence us away from our habits or choices.
    \item Overestimation bias: our semi-structured interviews showed an almost systematic overestimation of travel times by bike and while walking.
\end{itemize}

Following this study, a first Python simulation was implemented, that showed the interest of integrating cognitive biases in the decision process to enhance simulation realism. But before implementing cognitive biases in a complete urban mobility simulation, we need to more formally quantitatively evaluate their effect on mobility decisions. The survey designed to this aim is presented in the next section.

\section{Survey of mobility perceptions and priorities}   
\label{sec:survey}                                        

A quantitative analysis of the effect of biases on the choice of a daily mobility means requires a study on a large number of people. To ensure a large participation, we chose to use an online questionnaire form. The goal of this survey is to evaluate the rationality of mobility choice and to identify cognitive biases that could impact this choice.

\subsection{Methodology}

The questionnaire consists of three main parts. The first part concerns the responders' profile and their mobility habits. In the second part, the participants are asked to provide their ratings of priority over the 6 decision criteria identified above (ecology, comfort, financial accessibility, practicality, safety and speed), \ie to state how important each criterion is in their choice of a daily mobility mode. The third and last part concerns the participants' perceptions of the value of the mobility modes considered (bicycle, car, public transport and walking) over these criteria: they are asked to mark how well they think that each mode satisfies each criterion. All ratings (priorities and values) are made on a Likert scale from 0 to 10.

The goal of these questions is to allow us to estimate what would be the rational mobility choice of each participant with respect to the formula described above, based on the priorities and values that they provide. We should then be able to compare the declared mobility mode of each participant with that predicted rational mode. In the case when they differ, several reasons can explain the discrepancy, including the impact of cognitive biases. Besides, the declared values and priorities will be averaged to calibrate parameters of our simulator.

\subsection{Questionnaire}

The questionnaire was administered in French. Below is a translation of its main elements:\begin{footnotesize}
\texttt{
\begin{itemize}
    \item Part 1: responder profile 
    \begin{enumerate}
        \item What is your gender (Multiple choice: woman, man, other, do not wish to answer)
        \item Which mobility mode do you use most for your daily trips? (Multiple choice: bicycle, car, public transport, walking)
        \item What is the distance between your place of residence and your main place of activity, in km? 
        \item How many times per week do you realise the roundtrip between your place of residence and your main place of activity? 
        \item Are any mobility modes inaccessible to you (personal constraints, lack of infrastructures, for instance no public transport available)? (Multiple choice: bicyle, car, public transport, walking, none)
        \item Give precisions if you wish. (free text)
    \end{enumerate}
    \item Part 2: importance of choice criteria: for each criteria, we ask you to evaluate how important it is for you when choosing your daily mobility mode. Answers on a Likert scale from 0 to 10 for the following 6 criteria:
    \begin{itemize}
        \item Ecology: this mode has a low carbon footprint
        \item Comfort: this mode is pleasant to use
        \item Financial accessibility: this mode is cheap
        \item Praticity: this mode is flexible, does not impose strong constraints 
        \item Safety: this mode allows me to move without risks (accidents, injury, aggression)
        \item Speed: this mode allows me to reach my destination in low time
    \end{itemize}
    \item Part 3: evaluation of responder's perceptions of mobility means. For each of the 4 modes:
    \begin{enumerate}
        \item How would you rate the (ecology, comfort, financial accessibility, praticity, safety, speed) of this mode as a daily commuting mode? 6 questions, answers on a Likert scale from 0 to 10.
        \item Do you want to add anything on your perception of this mode? (free text answer)
    \end{enumerate}
\end{itemize}
}
\end{footnotesize}

\subsection{Participation}

\begin{wrapfigure}[7]{r}{0.65\textwidth}
    \centering
    \vspace*{-30pt}
    \includegraphics[scale=0.4]{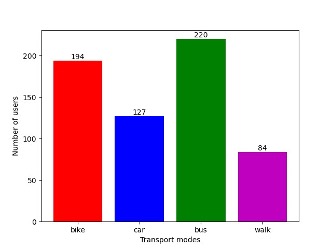}
    \caption{Distribution of habitual mobility mode over 625 respondents}
    \label{fig1}
\end{wrapfigure}

Following the diffusion of our survey online during spring 2023, 625 people did answer the questionnaire. Figure~\ref{fig1} represents the distribution of usual mobility mode in our respondents. We will use this distribution as a comparison point with our simulation results.

\bigskip\bigskip

\subsection{Average values to calibrate the simulator}

In order to calibrate our agent-based simulation, we will use the average values reported by the respondents for the priority of criteria, as well as for the evaluation of modes on these criteria.

\begin{figure*}[h]
    \centering
    \begin{subfigure}{0.23\textwidth}
        \includegraphics[scale=0.25]{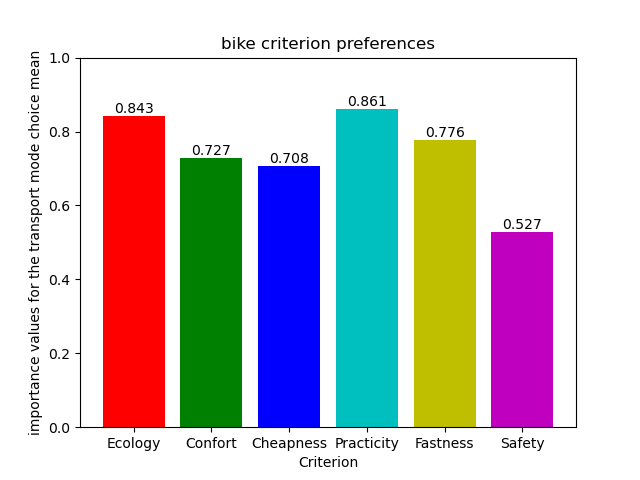}
        \caption{Cyclists}
    \end{subfigure}
    \begin{subfigure}{0.23\textwidth}    
        \includegraphics[scale=0.25]{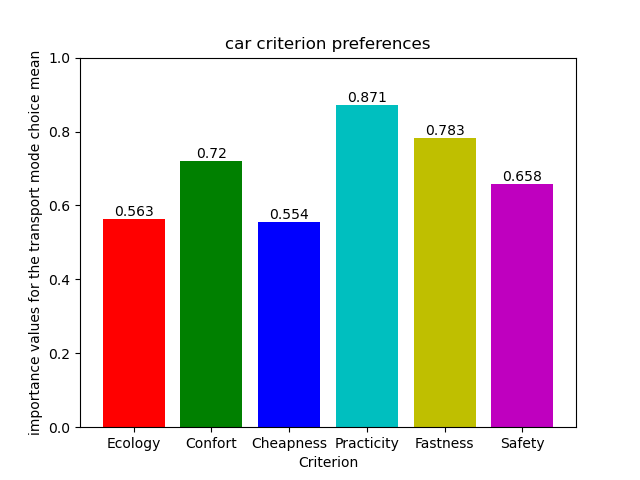}
        \caption{Car drivers}
    \end{subfigure}
    \begin{subfigure}{0.23\textwidth}    
        \includegraphics[scale=0.25]{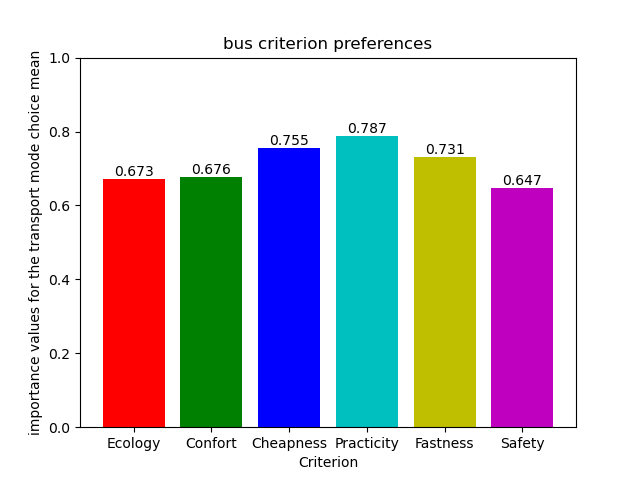}
        \caption{Bus users}
    \end{subfigure}
    \begin{subfigure}{0.23\textwidth}    
        \includegraphics[scale=0.25]{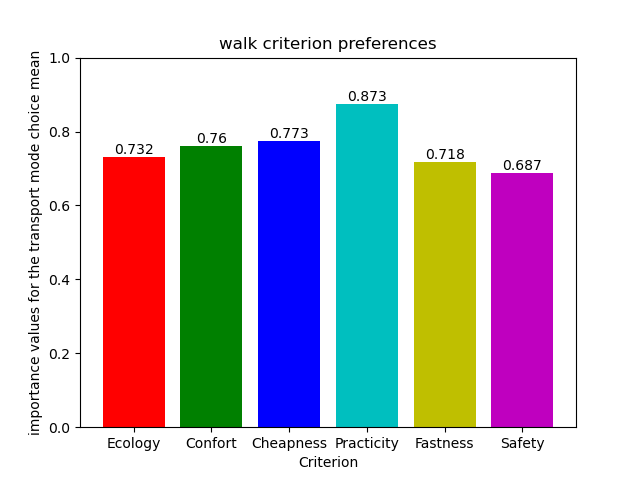}
        \caption{Walkers}
    \end{subfigure}
    \caption{Priority profiles of the 4 subpopulations wrt usual mobility mode}
    \label{fig14}
\end{figure*}

\paragraph{Average priorities of criteria.}
We first computed the average priority of each of the 6 decision criteria, over each subpopulation of respondents (cyclists, car drivers, bus users, pedestrians), as well as over the global population of all respondents. Figure~\ref{fig14} reports these average priorities, computed from the answers to the part 2 of our survey.

\begin{figure}
    \centering
    \begin{subfigure}{0.47\textwidth}
        \includegraphics[scale=0.25]{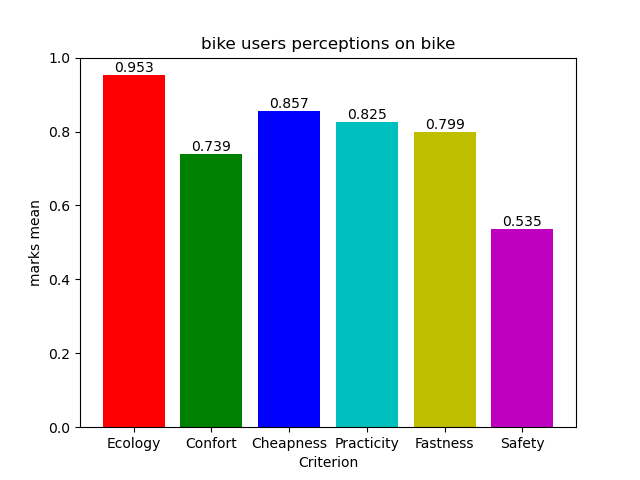}\includegraphics[scale=0.25]{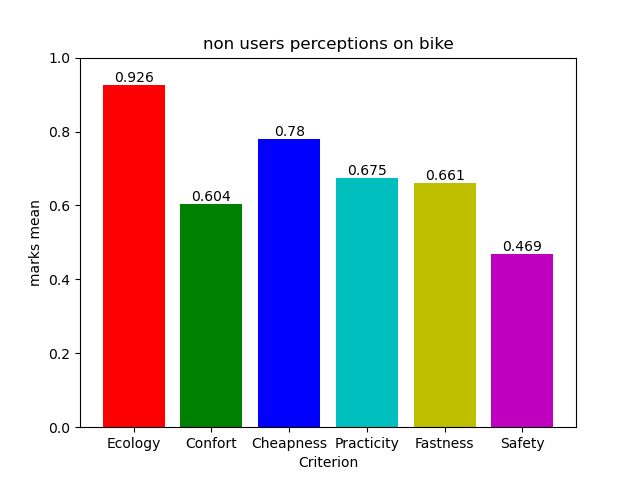}
        \caption{Bicycle}
    \end{subfigure}
    \hfill
    \begin{subfigure}{0.47\textwidth}
        \includegraphics[scale=0.25]{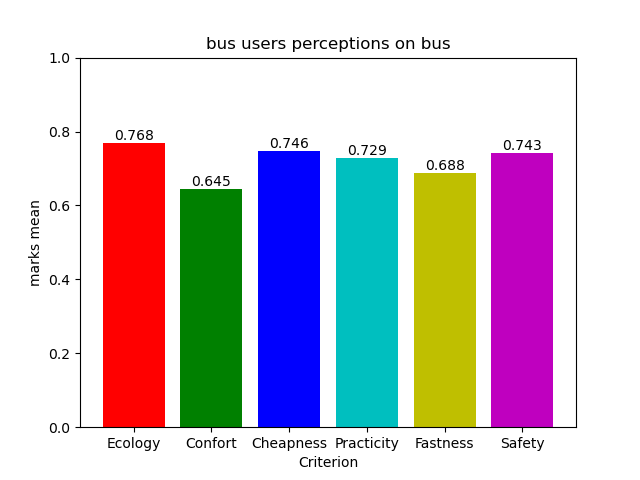}\includegraphics[scale=0.25]{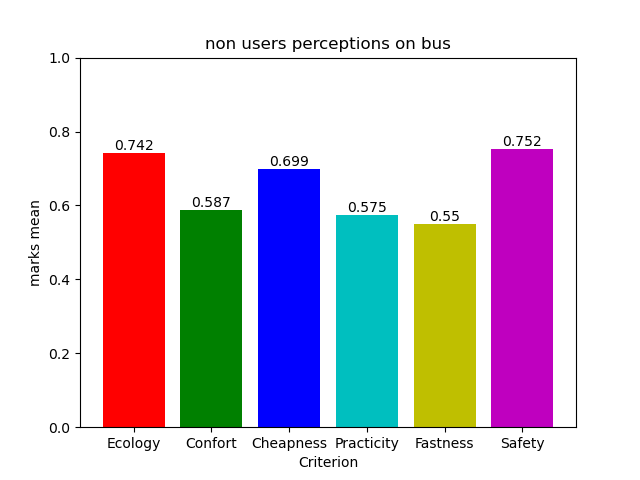}
        \caption{Bus}
    \end{subfigure} 
\hfill
    \begin{subfigure}{0.47\textwidth}
        \includegraphics[scale=0.25]{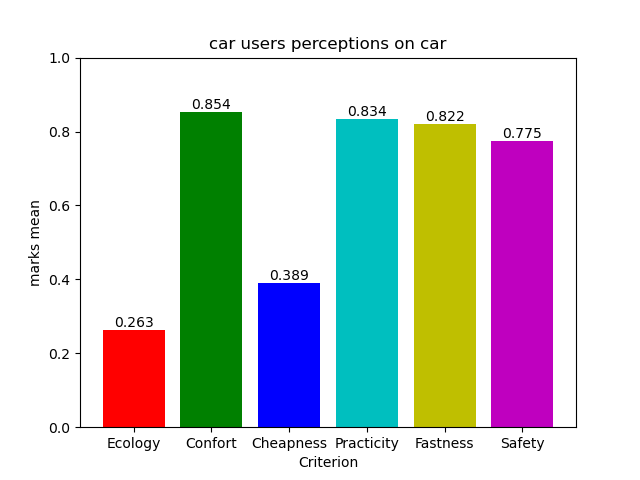}\includegraphics[scale=0.25]{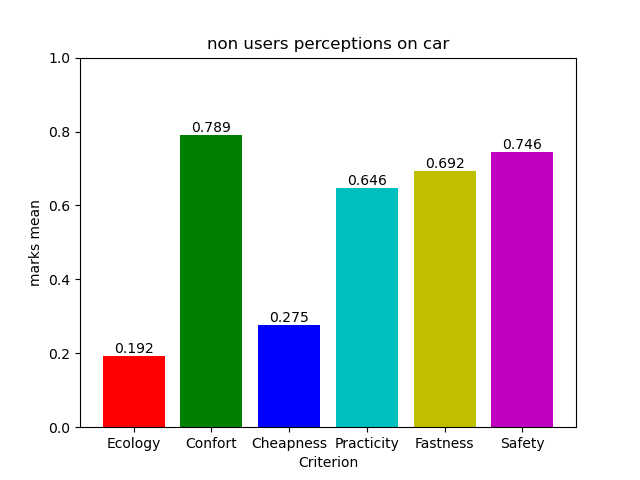}
        \caption{Car driving}
    \end{subfigure} 
\hfill 
    \begin{subfigure}{0.47\textwidth}
        \includegraphics[scale=0.25]{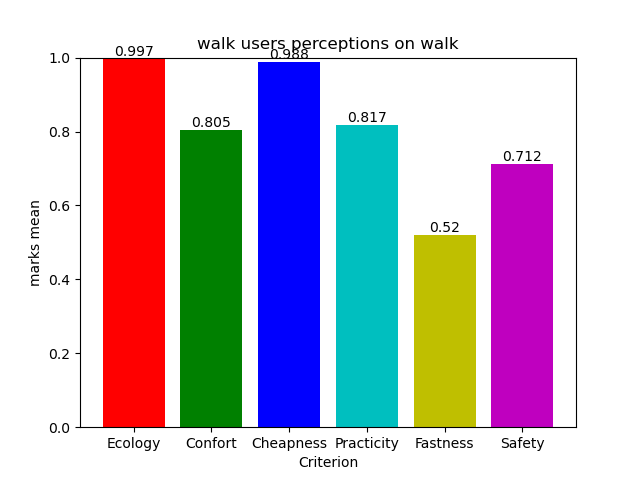}\includegraphics[scale=0.25]{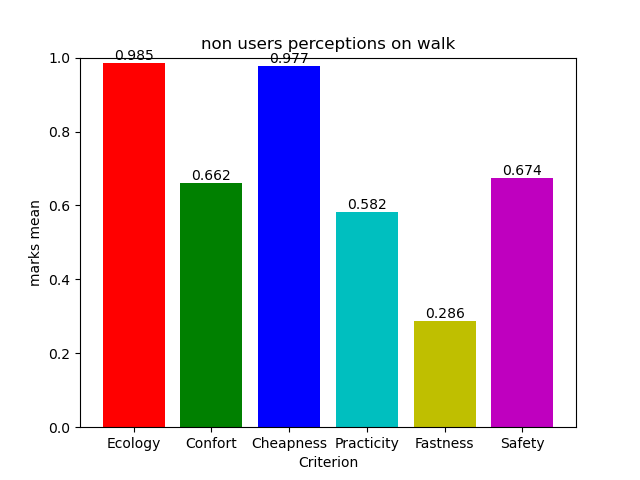}
        \caption{Walking}
    \end{subfigure} 
    \caption{Comparing the evaluation of modes by users / non-users}\label{fig15}
\end{figure}

\paragraph{Average values of modes on criteria}
We then computed the average evaluations of mobility modes on the 6 decision criteria, over the subpopulations of users of this mode, vs non-users (\ie users of all 3 other modes). This is based on the answers to the part 3 of our survey. Figure~\ref{fig15} reports the results.

We can see that the evaluation of the modes differ among their users vs the users of other modes. Concretely, people who use a mode rate it higher, on most criteria, than people who do not use it. This can be explained in two ways: people who rate a mode higher are therefore more likely to use it; but people who use a mode are also more likely to upgrade their perception \emph{a posteriori} to justify their choices.

\subsection{Rational choices and identification of biases}

The general idea of evaluating the rationality of respondents' choices of means of transport is to estimate, based on their responses, the choice that would be considered as rational by the multi-criteria decision model described above (Section~\ref{sec:soa}). We will then compare the predicted rational choice with the habitual mobility mode reported by the participant. We expect that not all respondents will use the mobility mode that would be considered rational, which can be explained by several reasons, such as individual constraints (the rational mode is not available or accessible to the user), habits (the user's habit is kept despite not being best anymore), or cognitive biases.

We cannot consider that a choice is irrational when it is the only one available. Therefore, as a first step, we eliminated from the list of possible choices the modes that the user ticked as unavailable (in part 1 of the survey) and the modes that would take too long based on the reported home-work distance (over 15km for bicycle, over 5km for walking). In order to estimate the best rational mobility choice for each participant, we then used two different methods.

\begin{figure*}[h]
    \centering
    \begin{subfigure}{0.24\textwidth}
        \includegraphics[scale=0.25]{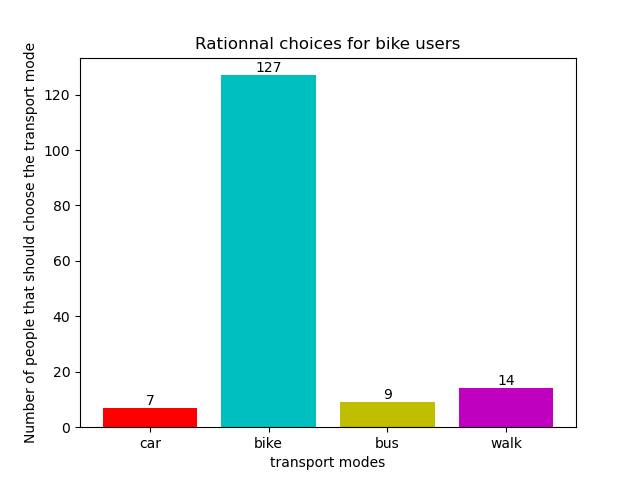}
        \caption{Cyclists}
    \end{subfigure}
    \hfill
    \begin{subfigure}{0.24\textwidth}
        \includegraphics[scale=0.25]{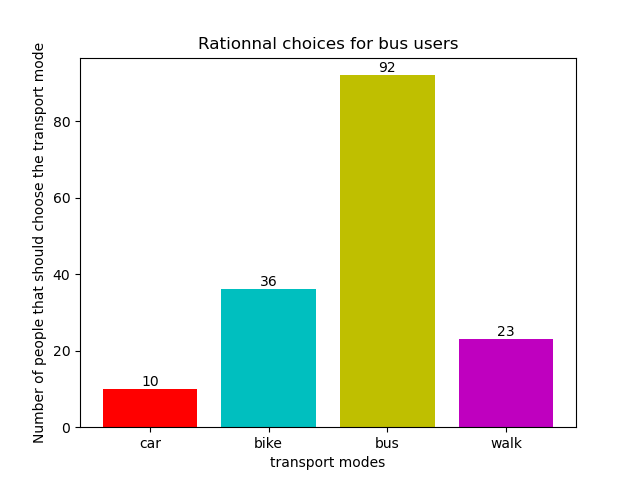}
        \caption{Bus users}
    \end{subfigure}
    \begin{subfigure}{0.24\textwidth}
        \includegraphics[scale=0.25]{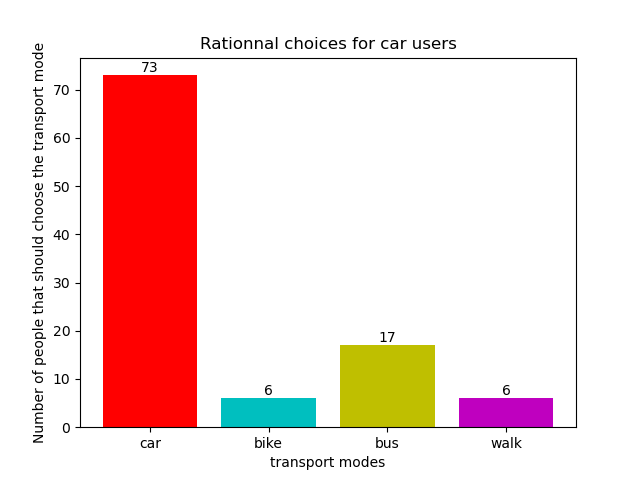}
        \caption{Car drivers}
    \end{subfigure}
    \begin{subfigure}{0.24\textwidth}
        \includegraphics[scale=0.25]{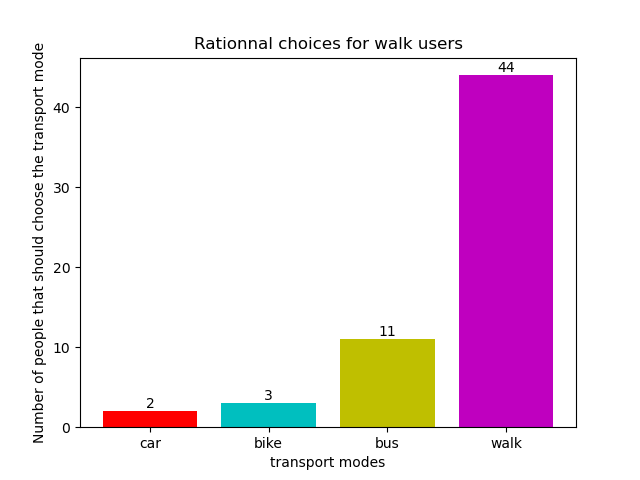}
        \caption{Walkers}
    \end{subfigure}
    \caption{Rational choice of mobility for users of the 4 modes, based on personal DECLARED values}
    \label{fig16}
\end{figure*}

\paragraph{Personal values.}
The first method consists in considering the evaluations of the modes on the six criteria that were reported by the respondents themselves (in part 3 of the questionnaire). These evaluations are used in the multi-criteria decision formula presented above, to compute the scores of all available modes; the mode with the best score is considered as the rational choice. This method allows us to identify participants who make 'irrational' choices with respect to their own perceptions of the modes. Figure~\ref{fig16} shows the resulting predicted rational choices of users of each mode, as computed with our multi-criteria decision formula, using their declared priorities and declared evaluations. We can observe that the chosen mode matches the rational prediction for many, but not all, users.

\paragraph{Crowdsourced values.}
However, these declared evaluations could themselves be biased, as suggested by the results above (Figure~\ref{fig15}). As a result, some participants could appear rational because of a biased reported evaluation. In order to solve this problem, we used a second prediction method. Since the value of modes on criteria depends mostly on the urban infrastructure (having safe cycling lanes, sufficient frequency and capacity of buses, expensive petrol and car insurance, etc) or on external parameters (bicycle is more ecological than car driving), we could consider them as homogeneous among all agents in a similar environment. We therefore decided to 'crowdsource' the evaluation of each mode on each criteria by computing its average evaluation over all respondents. We then used the same multi-criteria decision formula, but replacing the individually reported subjective evaluation, with the supposedly more objective, average value. Figure~\ref{fig16} reports the results. We can see that more choices are considered irrational with this second method, due to taking into account evaluation biases that were hidden by the first method.

\begin{figure*}[h]
    \centering
    \begin{subfigure}{0.24\textwidth}
        \includegraphics[scale=0.25]{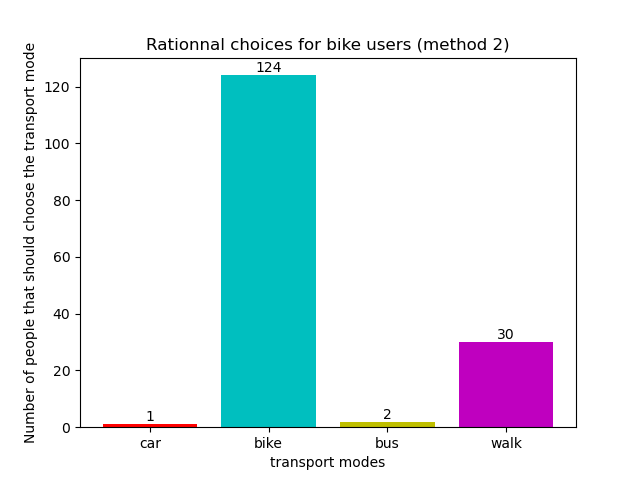}
        \caption{Cyclists}
    \end{subfigure}
    \hfill
    \begin{subfigure}{0.24\textwidth}
        \includegraphics[scale=0.25]{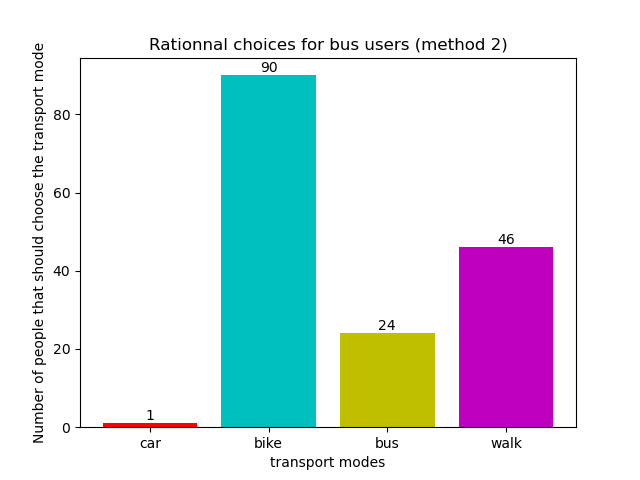}
        \caption{Bus users}
    \end{subfigure}
    \hfill
    \begin{subfigure}{0.24\textwidth}
        \includegraphics[scale=0.25]{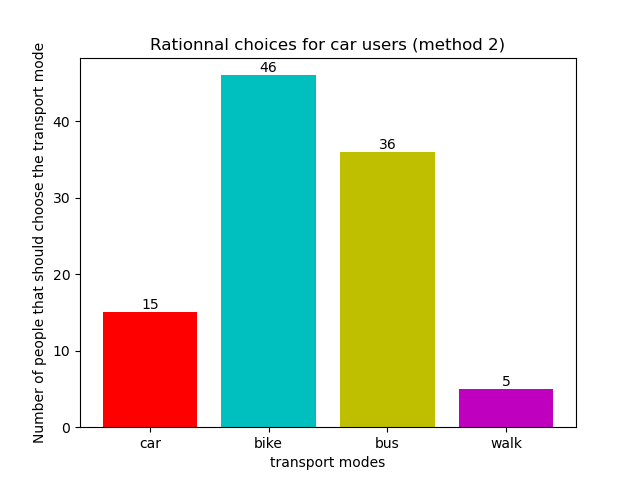}
        \caption{Car drivers}
    \end{subfigure}
    \hfill
    \begin{subfigure}{0.24\textwidth}
        \includegraphics[scale=0.25]{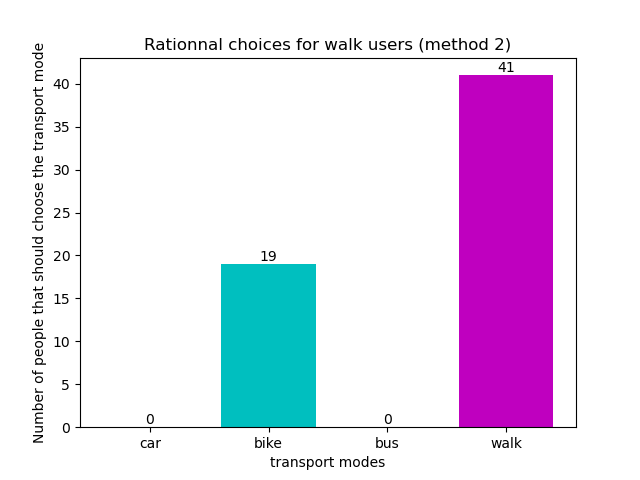}
        \caption{Walkers}
    \end{subfigure}
    \hfill
    \caption{Rational choice of mobility for users of the 4 modes, based on average CROWDSOURCED values}
    \label{fig17}
\end{figure*}

\subsection{Discussion}

First of all, it is interesting to observe the big gap between the results obtained by the two methods. Indeed, when using reported subjective evaluations in the formula (method 1), we find that 443 (out of 625) respondents make the rational choice of daily mobility. When using the average crowd evaluation in the same formula (method 2), we only have 297 rational choices among the respondents. 

This can be explained when studying the respondents' perceptions of the different mobility modes. Indeed, we can observe that the perception of the value of a mode on the criteria is over-evaluated by the users of this mode when compared with the general population. This is even more obvious when comparing with the perception by non-users of this mode (Figure~\ref{fig15}). Assuming that each mode has an objective value on each criteria (walking is perfectly ecological while driving is not, road safety can be evaluated by accident statistics, etc), we could explain this over-estimation by the impact of the confirmation bias. For instance, users of public transport could unconsciously modify their perception of this mobility mode to bring an internal validation of their choice. 

A similar bias can also be observed in the priority profiles of the respondents (Figure~\ref{fig14}). For instance, car drivers report a lower priority of ecology and financial accessibility than the other populations, while considering that individual car is the least ecological and the most expensive mode. Similarly, cyclists perceive the bicycle as being the most dangerous mode, but also report a lower priority for safety. Of course it is impossible to say if there is a selection of users who are already less sensitive to safety and therefore choose to ride a bicycle, or if people choosing to ride a bicycle will subsequently lower their priority for safety to validate their choice.

To summarize, we can observe that among the respondents to the questionnaire, only 47.5\% choose their daily mobility mode rationally according to the second method. One possible explanation for this phenomenon is the confirmation bias. But other biases could also have an effect without being identifiable from our results. For instance, reactance is hard to show in our results since it appears in response to an external incentive that is perceived as trying to force oneself into a different choice. Thus, to evaluate the effects of this bias, it would be necessary to voluntarily create this incentive and observe the evolution of perceptions on the different means of transport. Furthermore, the effects of overestimation bias could not be observed with the responses to the questionnaire; however it had already been clearly identified in the preliminary semi-structured interviews, and will thus be implemented in our simulation.

To conclude, it must also be clarified that the conclusions drawn from the responses to this questionnaire are only hypotheses for the development of a simulation. Evaluating the presence of cognitive biases in a decision using a questionnaire is complex, because it is possible that respondents adapted their answers to justify their choice of daily means of transport, and because several different reasons could explain the same decision. In the sequel, we implemented these hypotheses in an agent-based simulation in order to check whether they make it possible to generate results that are consistent with reality, which would give more value to this study.

\section{Agent-based mobility simulation}         
\label{sec:simul}                                 

Following the completion of this survey, we developed a multi-agent simulation of choices of daily mobility mode. It was decided to carry out this simulation in the Gaml language which is the programming language supported by the GAMA platform\footnote{GAMA platform: \url{https://gama-platform.org/}} \cite{gama19}. It is an open source multi-agent simulation platform with an integrated development environment. The Gaml language was created specifically for this platform, it was designed to be usable by non-computer scientists and is based on the Java language. Besides, it is already used in the previous versions of the Switch simulation so it was more coherent to continue the work with this same language.

\subsection{Agents}

An agent in the simulation represents an individual having to choose at each time step one mobility mode among the four ones considered.

\paragraph{Attributes and initialisation.}
Each agent has a number of attributes influencing its decisions. 
\begin{itemize}
    \item Physical fitness: float between 0 (very bad) and 100 (perfect).
    \item Home-work distance, randomly initialised between 0 and 20km, since we are mostly interested un urban mobility.
    \item Habits: the agent remembers which modes it used in the past; the history is randomly initialised and then updated with each trip. Habits are used to compite cognitive biases and to initialise the following 2 attributes
    \item Priorities for the 6 criteria: initialised with the results of our survey (average priorities over the subpopulation of users of a given mode), for the habitual mobility mode deduced from the agent's habits. These priorities are considered static.
    \item Perceptions of modes on criteria: also initialised from the results of our survey (average values of modes on criteria over the subpopulation of users of each mode) and based on the agent's habitual mode. These perceived values will dynamically evolve during the simulation, under the effect of cognitive biases or environmental changes.
\end{itemize}

\paragraph{Behaviour.}
At each time step of the simulation, all the agents choose a mode of mobility among the 4 ones considered (bicycle, car, public transport and walking), by computing their scores with the multi-criteria decision formula (weighed average of values of the mode on each criterion, weighed by the priority of the criterion). However, we modify this initial model by introducing several pre-processing operations on the agent's perceptions and the availability of means of transport.

Firstly, walking is removed from the options for agents with a physical fitness lower than 10 whatever the distance; for agents with a physical fitness between 10 and 90 if the distance is more than 2 km; and for agents with a physical fitness above 90 if the distance is above 5km. Bicycle is similarly removed for agents with physical fitness lower than 10; for agents with a home-work distance greater than 8 km and a physical fitness between 10 and 90; and for agents with a physical fitness over 90 and a distance over 20 km.

Second, the effects of cognitive biases are applied on perceived values of modes by some agents (the proportion of biased agents is a parameter of the simulation). For those agents that are concerned by the confirmation bias, the perceived values of their most used mobility mode on the criteria are raised. For the agents concerned with the overestimation bias, their perceived speed of walking and cycling are lowered.

\subsection{The environment.}
A central element of the simulation is the description of the environment in which the agents will evolve. This description is made using several quantitative (Table~\ref{tab:quanti}) and qualitative (Table~\ref{tab:quali}) parameters. Among these parameters, some will be directly taken into account in the behaviour of the agents while others represent a global description of the environment. A default value is assigned to each of these parameters. Most of these values were taken from the Python simulation developed earlier or from the existing version of the SWITCH simulation. Only the price of petrol was updated in January 2023, to correspond to the average price at this period.

\begin{table}[h]
    \centering
    \begin{tabular}{|c|c|}
    \hline
\textbf{Parameter} & \textbf{Default value} \\
\hline\hline
Number of agents & 625 \\
\hline
Petrol price & 1.95euro/L \\\hline
Monthly bus pass & 65.5 euros \\\hline
Proportion of roads with cycling lane & 50.0\%
\\\hline
Mean frequency of buses & 10 / hour
\\\hline
Mean bus capacity & 100 people 
\\\hline
Mean car speed & 42.3 km.h
\\\hline
Mean bicycle speed & 14 km.h 
\\\hline
Mean walking speed & 6.4 km.h
\\\hline
Mean bus speed & 10 km.h
\\\hline
Proportion of agents with confirmation bias & 50\%
\\\hline
Proportion of agents with over-estimation bias & 50\%
\\\hline
    \end{tabular}
    \caption{Quantitative parameters and their default values}
    \label{tab:quanti} 
\end{table}

\begin{table}[h]
    \centering
  \begin{tabular}{|c|c|}
\hline
\textbf{Parameter} & \textbf{Default value} \\
\hline\hline
Rainy weather & False\\\hline
Pleasant temperature & True \\\hline
Daylight & True \\\hline
Urban environment & False \\\hline
Rush hour & False \\\hline
\end{tabular}  
    \caption{Qualitative parameters and their default values}
    \label{tab:quali}  
\end{table}

\begin{figure}[h]
    \centering
    \includegraphics[scale=0.5]{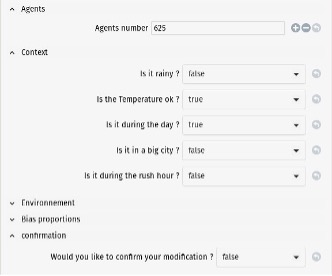}
    \caption{Screenshot of the graphical interface for modifying environment parameters of the simulation}
    \label{screen}
\end{figure}

The number of agents and the proportions of agents affected by the biases are global description parameters of the environment used at initialisation. The other parameters all have a direct effect on the choices of agents and can be modified during the simulation, via the graphical interface (Figure~\ref{screen}); the new values must be confirmed to notify the simulation of the changes, so that the agents' perceptions can be subsequently updated. If these parameters are left at their default value, they have no effect, the environment is considered neutral. If the user modifies their value, it impacts the agents' perceptions of the mobility modes concerned by the parameter. For example, if the user increases the proportion of roads equipped with cycle lanes, the safety and speed values of cycling increase. If the user activates the rainy weather setting, cycling becomes less comfortable and more dangerous, walking becomes less comfortable and driving becomes more dangerous.

\subsection{Simulation results}

We consider as output of our simulation the distribution of agents' choices over the four available mobility modes. Since we initialise our population with the priorities and values reported in our survey, we wish to obtain a resulting distribution similar to the one measured in our surveyed population (Figure~\ref{fig1}). We ran two experiments to compare the distributions obtained with and without the influence of cognitive biases.

\begin{figure}[ht]
    \centering
    \includegraphics[scale=0.5]{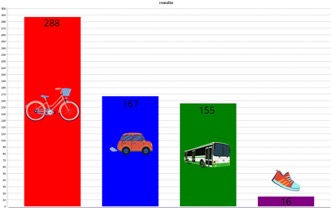}
    \caption{Mobility distribution resulting from experiment 1, without cognitive biases}
    \label{fig7}
\end{figure}

\newpage
\paragraph{Experiment 1: no cognitive biases.}
We ran a first experiment without activating cognitive biases (0\% of agents affected). Figure~\ref{fig7} shows the results. 

We can then see that this distribution is quite far from what is expected. First, the proportions of mobility modes are not correct: the number of cyclists is too high (288 versus around 194 expected) and the number of public transport users is too low (155 versus around 220 expected). Second, the ranking order of the modes relative to each other is not even respected: public transport should come first (according to our surveyed population) but is only third in this simulation. For the simulation to be considered as realistic, it is necessary that we can at least reproduce the observed rankings.

\begin{figure}[ht]
    \centering
    \includegraphics[scale=0.5]{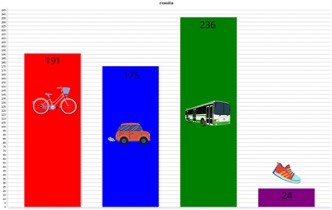}
    \caption{Results of experiment 2: distribution of mobility with cognitive biases}
    \label{fig8}
\end{figure}

\paragraph{Experiment 2: with cognitive biases.}
Figure~\ref{fig8} shows the results of a second experiment, where cognitive biases were activated. These results are closer to what is expected. The ranking of the modes by proportion of users is respected. However, the number of car drivers remains a little too high, while the number of walkers is too low.

\section{Conclusion} \label{sec:cci}   

The processing of the questionnaire data clearly showed that the choice of a means of daily transport is not always made rationally. We were also able to show that the confirmation bias could play a role in the irrationality of this choice. This questionnaire associated with previous work in the project allowed to design and implement an agent-based simulation taking into account confirmation bias and overestimation bias in the agents' decisions.

In view of our results, the implementation of cognitive biases in an agent-based simulation seems to be an interesting avenue to make it more realistic. However, for the results to be better, these biases should be better formalized so that they function as in our cognitive system. For this, it will be necessary to carry out in-depth work in cognitive sciences to understand their real functioning.

In addition, one way to improve our simulation is to implement a form of interaction between the agents, which is not yet taken into account in our current model. Indeed, some cognitive biases such as reactance emerge from interactions between agents, and decisions are also influenced by social biases (social pressure, sheep effect). Future work is therefore needed to improve the simulation with such social aspects.

\section*{Acknowledgements}

This internship was co-supervised by Marie LEFEVRE from Univ. Claude Bernard Lyon 1, LIRIS lab. This work is part of the ANR project SwITCh funded by the French National Research Agency under number ANR-19-CE22-0003.

\bibliographystyle{abbrv}

\vfill
\section*{Appendix}

In this annex, we report the detailed average perceptions of how well each mobility mode satisfies each criteria, over different populations: the general population, and the subpopulation of users of each mode. These are the values used to initialise the agent population in the simulation.

\begin{figure}
    \centering
    \begin{subfigure}{\textwidth}
        \includegraphics[scale=0.25]{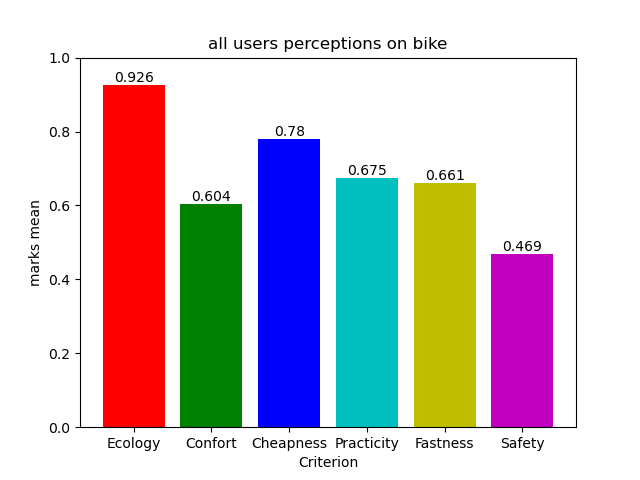}
        \includegraphics[scale=0.25]{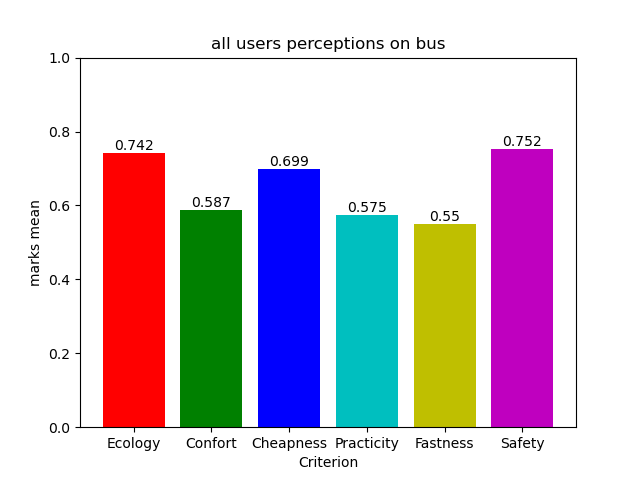}
        \includegraphics[scale=0.25]{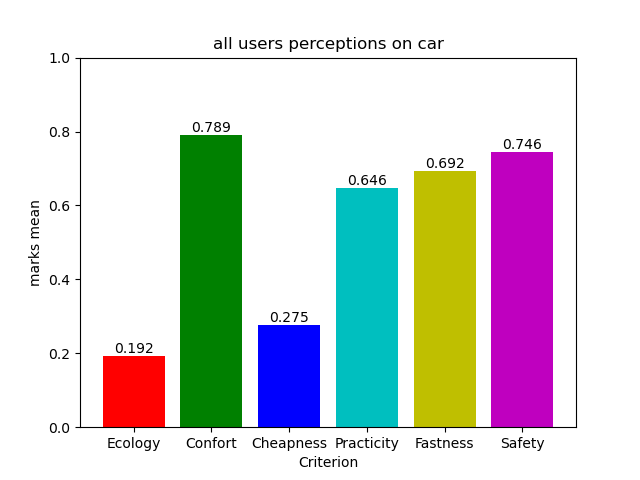}
        \includegraphics[scale=0.25]{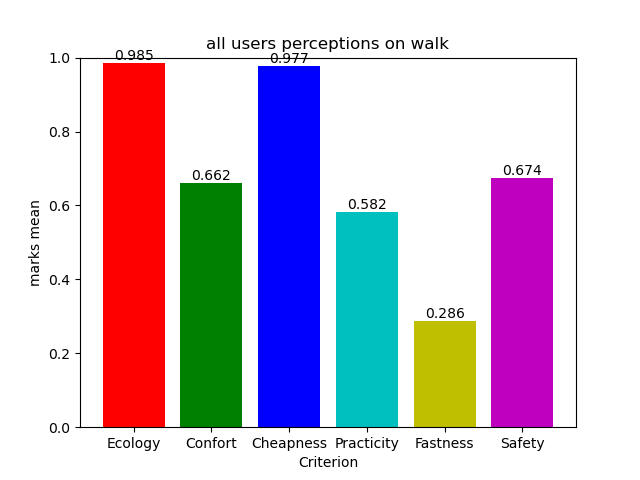}
        \caption{Global population of 625 respondents}
    \end{subfigure}
    \begin{subfigure}{\textwidth}
        \includegraphics[scale=0.25]{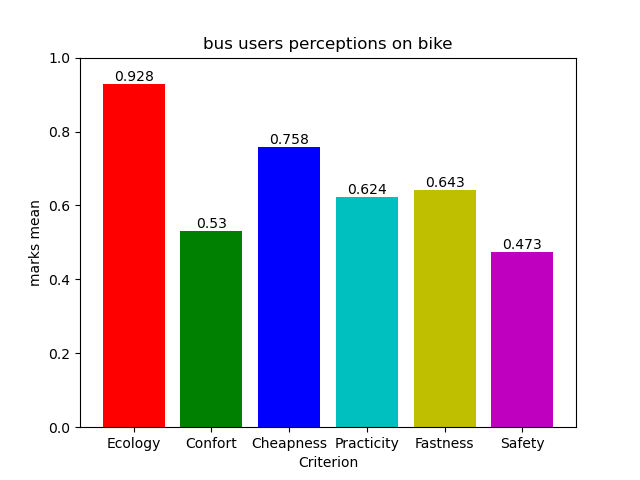}
        \includegraphics[scale=0.25]{imgs/bususers_perceptions_bus.png}
        \includegraphics[scale=0.25]{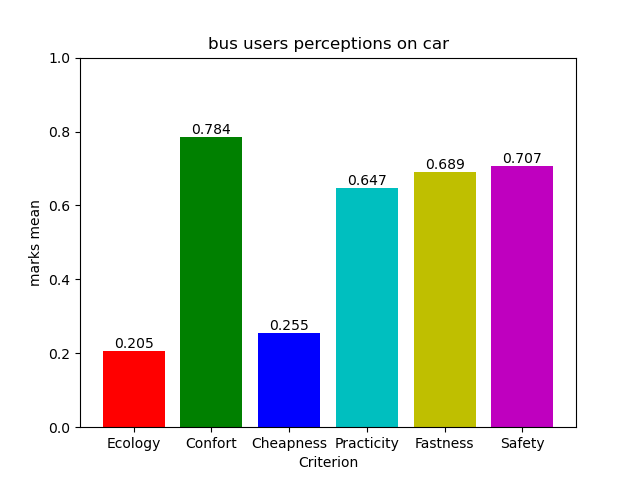}
        \includegraphics[scale=0.25]{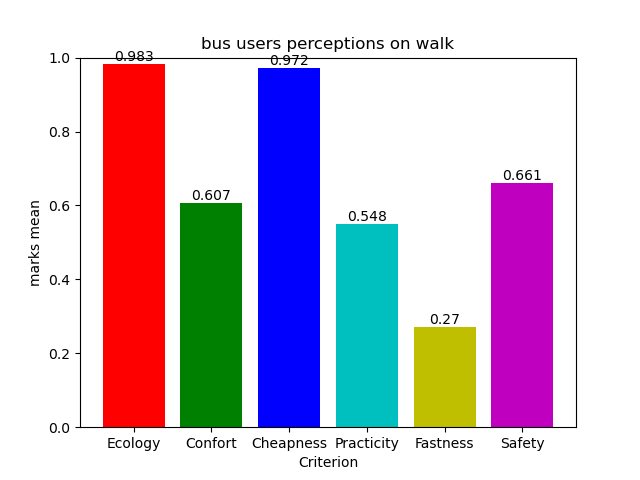}
        \caption{Bus users}
    \end{subfigure}
    \begin{subfigure}{\textwidth}
        \includegraphics[scale=0.25]{imgs/bike_perceptions_bike.png}
        \includegraphics[scale=0.25]{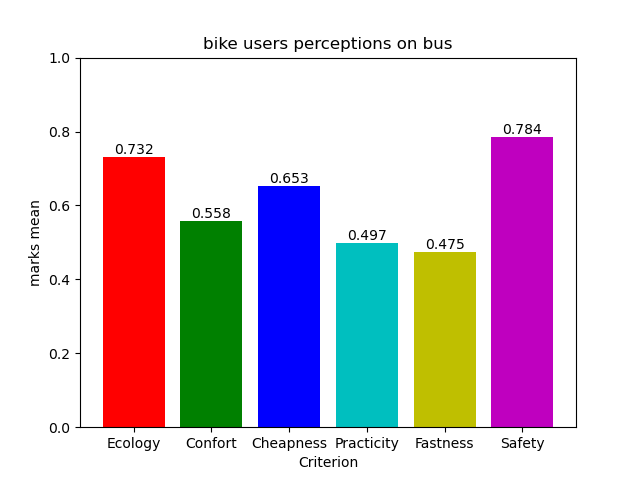}
        \includegraphics[scale=0.25]{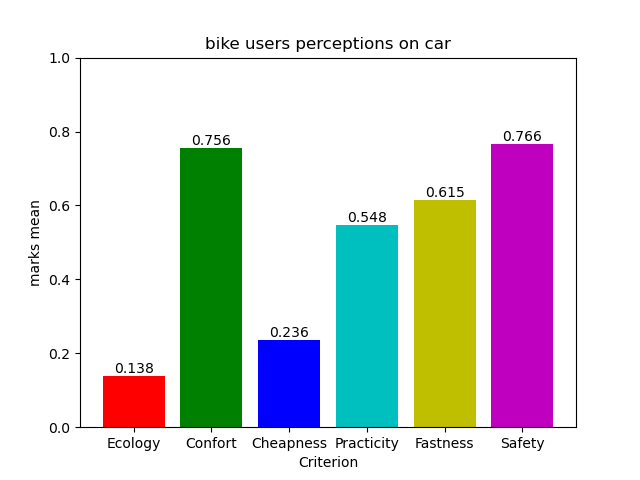}
        \includegraphics[scale=0.25]{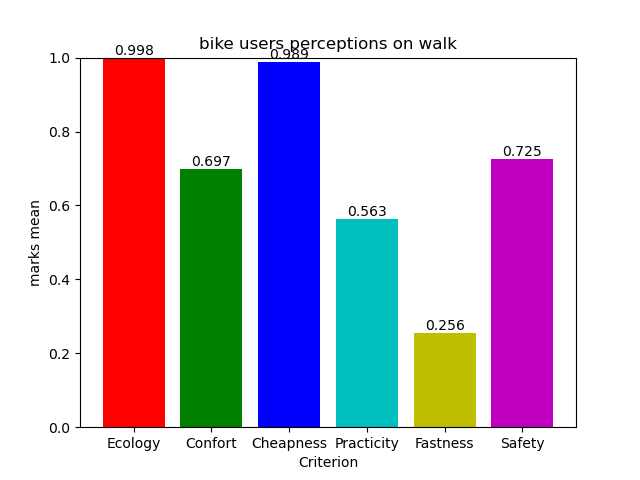}
        \caption{Cyclists}
    \end{subfigure}
    \begin{subfigure}{\textwidth}
        \includegraphics[scale=0.25]{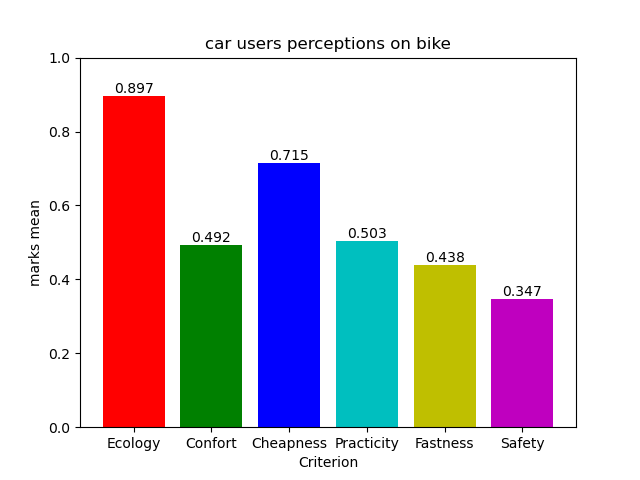}
        \includegraphics[scale=0.25]{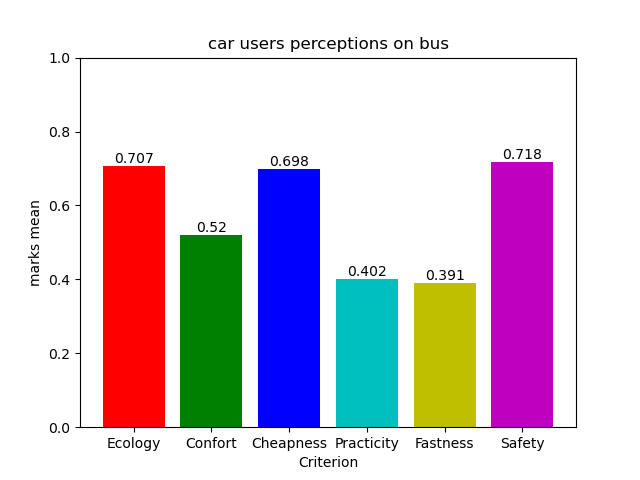}
        \includegraphics[scale=0.25]{imgs/carusers_perceptions_car.png}
        \includegraphics[scale=0.25]{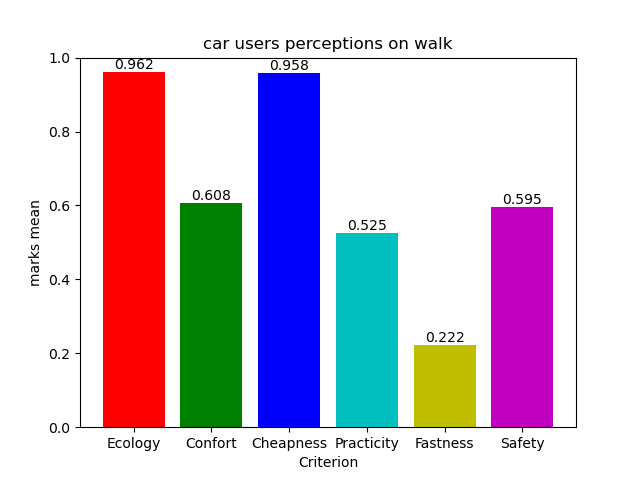}
        \caption{Car drivers}
    \end{subfigure}
    \begin{subfigure}{\textwidth}
        \includegraphics[scale=0.25]{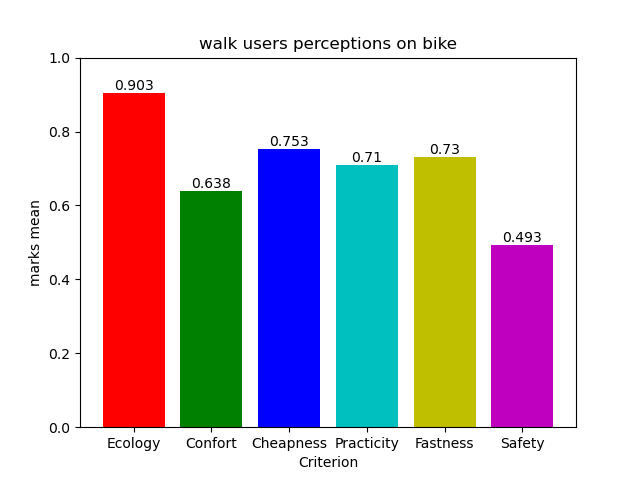}
        \includegraphics[scale=0.25]{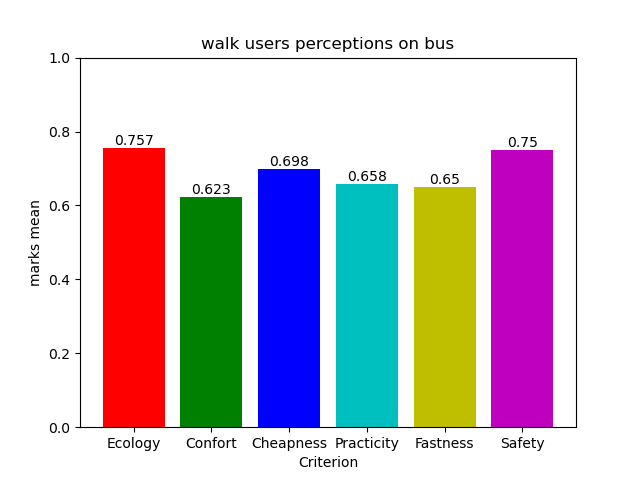}
        \includegraphics[scale=0.25]{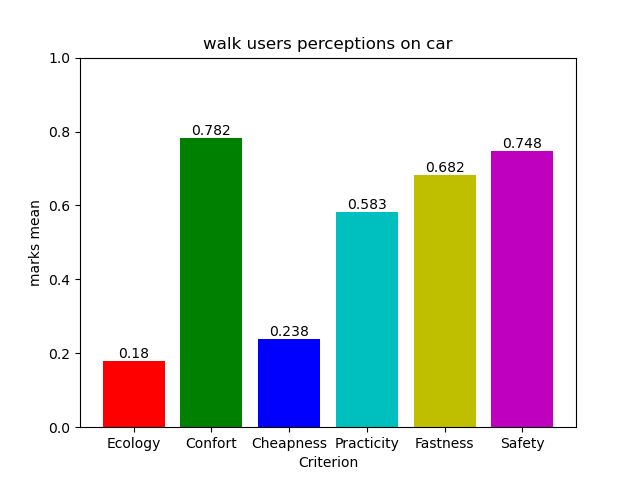}
        \includegraphics[scale=0.25]{imgs/walkers_perceptions_walk.png}
        \caption{Pedestrians}
    \end{subfigure}

    \caption{Comparing the average evaluations of the 4 modes, by subpopulation}
    \label{fig:annexa}
\end{figure}

\end{document}